\documentclass[twocolumn,showpacs,preprintnumbers,amsmath,amssymb]{revtex4}
\usepackage{graphicx}
\usepackage{epsfig}  
\usepackage{epsf}    
\usepackage{dcolumn}
\usepackage{bm}

\def\issue(#1,#2,#3){{\bf #1}, #2 (#3)} 

\def\opcit(#1){ {\em op. cit.}, #1}
\def\etal {\em et al.}

\def\APP(#1,#2,#3){Acta Phys.\ Polon.\ \issue(#1,#2,#3)}
\def\ARNPS(#1,#2,#3){Ann.\ Rev.\ Nucl.\ Part.\ Sci.\ \issue(#1,#2,#3)}
\def\CPC(#1,#2,#3){Comp.\ Phys.\ Comm.\ \issue(#1,#2,#3)}
\def\CIP(#1,#2,#3){Comput.\ Phys.\ \issue(#1,#2,#3)}
\def\EPJC(#1,#2,#3){Eur.\ Phys.\ J.\ C\ \issue(#1,#2,#3)}
\def\EPJD(#1,#2,#3){Eur.\ Phys.\ J. Direct\ C\ \issue(#1,#2,#3)}
\def\IEEETNS(#1,#2,#3){IEEE Trans.\ Nucl.\ Sci.\ \issue(#1,#2,#3)}
\def\IJMP(#1,#2,#3){Int.\ J.\ Mod.\ Phys. \issue(#1,#2,#3)}
\def\JHEP(#1,#2,#3){J.\ High Energy Physics \issue(#1,#2,#3)}
\def\MPL(#1,#2,#3){Mod.\ Phys.\ Lett.\ \issue(#1,#2,#3)}
\def\NP(#1,#2,#3){Nucl.\ Phys.\ \issue(#1,#2,#3)}
\def\NIM(#1,#2,#3){Nucl.\ Instrum.\ Meth.\ \issue(#1,#2,#3)}
\def\PL(#1,#2,#3){Phys.\ Lett.\ \issue(#1,#2,#3)}
\def\PRD(#1,#2,#3){Phys.\ Rev.\ D \issue(#1,#2,#3)}
\def\PRL(#1,#2,#3){Phys.\ Rev.\ Lett.\ \issue(#1,#2,#3)}
\def\SJNP(#1,#2,#3){Sov.\ J. Nucl.\ Phys.\ \issue(#1,#2,#3)}
\def\ZPC(#1,#2,#3){Zeit.\ Phys.\ C \issue(#1,#2,#3)}

\def\bra {\langle}
\def\ket {\rangle}

\def\l {\lambda}

\def\bl {\bar{\lambda}}

\def\d {\delta}

\def\bar {\overline}
\def\bbbar {B^0-\bar{B^0}}

\def\bpipi {B\to\pi\pi}
\def\bpik  {B\to \pi K}
\def\acp {A_{CP}}

\def\psiks {J/\psi K_S}

\def\be {\begin{equation}}
\def\ee {\end{equation}}
\def\bea {\begin{eqnarray}}
\def\eea {\end{eqnarray}}

\def\bc {\begin{center}}
\def\ec {\end{center}}

\begin{document}

\preprint{CU-PHYSICS-18/2004}
\preprint{hep-ph/0407061}

\title{Large electroweak penguins in $B\to\pi\pi$ and $B\to\pi K$ : 
implication for new physics}

\author{Soumitra Nandi and Anirban Kundu}
\affiliation{Department of Physics, University of Calcutta,\\
92 A.P.C. Road, Kolkata 700009, India}

\date{\today}

\begin{abstract}

We show that the anomalies in the $\bpipi$ decay channels, {\em viz.}, the
large branching ratio of $B\to\pi^0\pi^0$ and large direct CP asymmetry of
$B\to\pi^+\pi^-$, can both be explained, even in the framework of existing 
theoretical models which predict the relative strengths of the tree and the
strong penguin amplitudes, if there is an abnormally large electroweak 
penguin contribution. We also critically examine the expectation of a 
similarly large electroweak penguin in the $\bpik$ sector, and show that
it can be accomodated, and may even be motivated from an SU(3) flavor
symmetry between the $\bpipi$ and the $\bpik$ amplitudes, but is not 
necessary to explain the data. We emphasize that if the experimental numbers 
on these charmless nonleptonic channels survive, they may turn out to be an 
indirect signal of new physics. Current data is insufficient to discriminate 
betwen flavor-specific and flavor-blind new physics scenarios.

\end{abstract}


\maketitle

\section{Introduction}

Nonleptonic B meson decays pose a serious theoretical challenge to us.
The main difficulty is to handle the uncertainties of low-energy QCD; to
be very precise, to calculate the long-distance contribution $\bra M_1M_2|
{\cal H}_{eff}|B\ket$.  
There are at least three models with varied degrees of sophistication to
tackle the said uncertainty: the conventional factorization (CF) model
\cite{bsw,ali}, the QCD-improved factorization (QCDF) model \cite{beneke1,
beneke2}
and the perturbative QCD (PQCD) model \cite{keum1,keum2}. These models provide
definite theoretical predictions for branching ratios (BR) and CP
asymmetries ($\acp$) for nonleptonic charmless B decay modes. 
We still do not have
comparable numbers from other approaches, {\em e.g.}, the soft-collinear
effective theory (SCET) model \cite{bauer}, though they have predictions
for some of the $\bpipi$ amplitudes.  

One of the major obstacles is to calculate the strong phase, and hence the
CP asymmetry, theoretically. (The direct CP asymmetry depends both on
the strong phase difference and the weak phase difference of the two
interfering amplitudes, and is zero if any one of them vanishes.) 
The color-transparency argument of Bjorken \cite{bjorken} is not expected
to hold beyond the color-allowed tree decays. In the CF model the strong
phases are computed from the imaginary parts of the respective Wilson
coefficients \cite{ali}, and no soft final-state interactions are taken
into account. Among the more sophisticated versions, 
QCDF predicts a rather small strong phase difference between the dominant
amplitudes. PQCD 
predictions are different as annihilation and exchange topologies are given
more weightage than in QCDF (and the penguin amplitude is also enhanced),
and they can generate a sizable strong phase. (A comparative study of these
two models can be found in \cite{beneke2}). This is why
the predictions for CP asymmetries in QCDF and PQCD are sometimes even
opposite in sign. Needless to say, the relative importance of
different amplitudes varies from model to model. In particular, strong penguins
are much larger in the PQCD model due to a dynamical enhancement.
There are, however, some
common trends, which can be justified even intuitively. For example, the
tree-amplitude in $B\to\pi^+\pi^-$ dominates over the strong penguin, whereas
the situation is opposite for $B\to\pi^-K^+$. This is entirely due to
the relative importance of the CKM matrix elements. 

All these models also agree on the fact that the electroweak penguins (EWP)
are much smaller than the strong penguins. That can be seen very easily, by
comparing the relevant Wilson coefficients. Intuitively, it follows from
the smallness of $\alpha$ compared to the strong coupling constant $\alpha_s$. 
In fact, one can safely 
neglect all the electroweak penguins except that mediated by a top quark
in the loop. However, one must remember that the strong phases coming
from gluonic penguins (which is necessarily $I=0$) need not be the same
as those coming from $\gamma,Z$ penguins (they do not have a definite
isospin). Again, this complication does not matter since the EWPs are supposed
to be very small. There is no known mechanism which can enhance the EWP 
amplitude to the level of, say, the corresponding gluonic penguins. 

The predictions from the theoretical models are supposed to be robust when 
both the daughter mesons are light. For example, BRs and $\acp$s have been
computed for all the $\bpipi$ and $\bpik$ modes. Such numbers mostly agree
with each other, if we take into account the theoretical uncertainties.
This leads us to believe that the leading predictions of these models may
be taken seriously. These include the relative importance of the tree and
strong penguin amplitudes, the prediction for strong phase differences,
et cetera.

However, as is well known and will be substantiated in the next section,
some of the experimental numbers are at variance with the theoretical
predictions, though most of them agree. In the $\bpipi$ sector, there are
three such discrepancies:
(i) $Br(B\to\pi^0\pi^0)$ is abnormally
large (the largest prediction is from PQCD, which is about 5 times less);
(ii) $Br(B\to\pi^+\pi^-)$ is slightly on the lower side (the lowest
prediction is again from PQCD, which is more than $2\sigma$
above the data; QCDF has a large theoretical uncertainty \cite{beneke2}, 
and taking everything into account, may just fit the data); 
and (iii) the direct CP asymmetry in $B\to\pi^+\pi^-$
is larger than theoretical prediction, irrespective of the model
chosen. One must be cautious: the error bars, particularly for the
CP asymmetry data, are still large, and it may be too early to draw 
any definite conclusions based on them. Still, the trend is worth
investigating. 
  
In contrast, the individual BRs and $\acp$s in the $\bpik$ sector are
more or less in agreement with the theory. But when one eliminates the
theoretical uncertainties by taking suitable ratios of the BRs, it is
claimed \cite{buras1,buras2,hiller} 
that the EWP contribution is abnormally large.
It is more puzzling since it appears that we understand the
radiative decay $b\to s\gamma$ fairly well (though there are ways to
evade this problem). We will critically investigate
this issue also.
It will be shown that solutions with vanishingly small EWP amplitudes
can be obtained even with a rough SU(3) flavor symmetry between the 
tree and the strong penguin amplitudes of $\bpipi$ and $\bpik$ channels.
The strong phase difference between these two amplitudes for $\bpik$ 
is small (modulo $\pi$), but is different from the analogous quantity
for $\bpipi$ decays, which is exactly what is predicted in PQCD.
However, a substantially large EWP can be easily accomodated in the data.  
But there is no apparent conflict between the $\bpik$ channels
and the radiative decay $b\to s\gamma$.

There are two ways of analyzing the data. First, one can take a particular
model of his or her choice, and fit the data to get an idea of the relatively
poorly known parameters. The trouble is that equal justice cannot be done
to all the experiments. (A well-known example is the pre-2003
CKMfitter fit of the
$B\to\pi^+\pi^-$ asymmetry using the BaBar data only, within the context
of QCDF; the Belle data was so far away that it could not be fitted.) The
second approach \cite{buras1, gronau} 
is to analyze the data without paying any heed to the 
models, except, maybe, for some basic symmetry ideas like flavor SU(3). 
As expected, the fitted amplitudes and strong phases, all a priori
unknown, come out to be different from the model predictions.

The question is whether these discrepancies are due to the fact that we
do not understand the low-energy QCD well, or whether there is new physics 
(NP) beyond the Standard Model (SM). If there is NP, that will generate 
some more effective four-Fermi operators, leading to new contributions
in B decays. If these operators are flavor-blind, they should mimic the
strong penguins, whereas the flavor-specific operators should appear more
like a modification to the tree or the EWP amplitudes. The exact structure,
of course, will depend on the specific model.
%
Unfortunately, due to lesser number of $\bpipi$ 
channels, it is not possible to extract 
an observable which will clearly point to some abnormality in the EWP
sector, and we have to look for an indirect answer. 

The large BR of the $\pi^0\pi^0$ channel tells that if unknown EWP
dynamics is the cause of the enhancement, the EWP amplitude should
be quite large (we will later see that it is almost as large as the
color-allowed tree amplitude). It is difficult to have such a large
amplitude from intuitive arguments. The same may be said for the
$\bpik$ channels. (That is why we prefer a NP solution for the puzzle.) 

This brings us to the justification of our approach. We assume that the 
model predictions are sensible, as far as the tree and strong penguins (and
annihilation and exchange diagrams) go, and as a typical example we take
the PQCD model. Qualitatively same results are found for other models too,
but there is a reason why we choose PQCD. As we will show, a fit to the
$\bpipi$ data needs large penguin contribution, and only PQCD can provide
such a large enhancement, so that this is a more conservative approach to
look for NP effects. Also, a large $|P/T|$ helps the angle $\gamma$ to be
fitted in the first quadrant, in accordance with the standard CKM fit
\cite{ckmfitter}.  
We treat the parameters of the EWP sector as unknowns, and fit them 
from experimental data. This will show us that it is the 
$\pi\pi$ sector, rather than $\pi K$, that contains the puzzling EWP behaviour; 
and one should expect some deviation from the SM prediction for $B\to \rho
\gamma$, instead of $B\to K^*\gamma$.
We will also discuss how a large EWP amplitude can be correlated between
$\bpipi$ and $\bpik$ modes.

However, the main emphasis lies elsewhere. Such an analysis is the 
first step to look for effects coming from NP, since 
a number of NP models mimic EWP dynamics, by having unequal strength of
$b\to u\bar{u}d$ and $b\to d\bar{d}d$ amplitudes. There are numerous examples,
starting from models with an extra $Z'$ \cite{new-z} and/or
exotic quarks ({\em e.g.}, vector singlets) to R-parity conserving and 
violating supersymmetry \cite{rpc-rpv}. 
We do not go into any of these specific models; this is
a model-independent study to show that such models may have an interesting
future.

We wish to point out that this study implicitly assumes the unitarity of a
$3\times 3$ CKM matrix (thus, allowed ranges for the amplitudes for a 
model with a fourth chiral generation should be different). A weaker 
assumption is no NP effect in $\bbbar$ mixing. Such an effect modifies the
phase of the $\bbbar$ box, and $\beta$ as measured from $B\to\psiks$ need
not be the true $\beta$. This also weakens the limits on $V_{td}$. Inclusion
of such NP effects in $\bbbar$ mixing does not invalidate our conclusions;
it only changes the allowed range of the angle $\gamma$ of the Unitarity
Triangle (UT).   

The paper is arranged as follows. In Section II, we tabulate the experimental
and theoretical inputs for the analysis. Section III and IV deal with the 
$\bpipi$ and $\bpik$ data respectively. In Section V, we summarize and
conclude.

\section{Theoretical and Experimental Inputs} 

Let us first set our notations and conventions. We work in the
($\alpha,\beta,\gamma$) convention of the UT. 
We use $B$ to indicate a flavor-untagged $B^0$ or $\bar{B^0}$, or even a 
charged B meson when there is no chance for confusion.

The CP asymmetry for $B^+\to f$ is defined as
\begin{equation}
\acp = {\Gamma(B^+\to f) - \Gamma(B^-\to \bar f)\over
\Gamma(B^+\to f) + \Gamma(B^-\to \bar f)},
\end{equation}
which agrees with the convention of \cite{ali,beneke1,beneke2} but 
is opposite to that used by \cite{keum1,hfag}.
The same convention 
(according to the quark content, $b$ or $\bar{b}$) is
used for neutral B decays to flavor-specific final states, like $\pi^\pm
K^\mp$. For a flavor-nonspecific final state $f$ ({\em e.g.}, $\pi^+\pi^-$),
we define 
$\l=\exp(-i2\beta)\bra f|{\cal H}|\bar{B^0}\ket / 
\bra f|{\cal H}|B^0\ket$, and $a^d_{CP}=(1-|\l|^2)/(1+|\l|^2)$,
$a^m_{CP}=2{\rm Im}\l/(1+|\l|^2)$.
For the $\pi^+\pi^-$ system, they are related to the $S_{\pi\pi}$ and 
$C(A)_{\pi\pi}$ of BaBar and Belle by
\begin{equation}
S_{\pi\pi}=-a^m_{\pi\pi}, \ C_{\pi\pi}=-A_{\pi\pi}=a^d_{\pi\pi}.
\end{equation}

We also use the following valence quark convention:
\begin{eqnarray}
&{}&B^0\equiv \bar{b}d, B^+\equiv \bar{b}u, K^0\equiv \bar{s}d,\nonumber\\
&{}&K^+\equiv \bar{s}u, \pi^+\equiv u\bar{d}, \pi^0\equiv (u\bar{u}-d\bar{d})/
\sqrt{2},
\end{eqnarray}
and for the antiparticles, the quark contents are just reversed, without
any additional $-$ sign. This is different from those used by the
Gronau-Rosner group \cite{gronau}, but is consistent with, say, \cite{ali}. 
In the convention of \cite{gronau}, the following mesons have an extra
minus sign in their wavefunctions: $B^-,K^-,\pi^0,\pi^-$. 
This obviously does not change the BRs or $\acp$s. However, they change the 
amplitude sum rules. This will be discussed later in this section.

The experimental data (updated for Winter 2004), taken from the HFAG website 
\cite{hfag} and quoted at 68\% confidence level (CL), is as follows.

The BRs, multiplied by $10^6$, are
\begin{eqnarray}
Br(\pi^+\pi^-) &=& 4.6\pm 0.4;\nonumber\\
Br(\pi^0\pi^0) &=& 1.9\pm 0.5;\nonumber\\
Br(\pi^\pm\pi^0) &=& 5.2\pm 0.8;\nonumber\\
Br(K^0\pi^+) &=& 21.8\pm 1.4;\nonumber\\
Br(K^+\pi^0)&=& 12.8^{+1.1}_{-1.0}; \nonumber\\
Br(K^+\pi^-) &=& 18.2\pm 0.8;\nonumber\\
Br(K^0\pi^0) &=& 11.9^{+1.5}_{-1.4}.
\end{eqnarray}
The CP asymmetries are
\begin{eqnarray}
a^d_{\pi\pi} = C_{\pi\pi} &=& -0.46\pm 0.13;\nonumber\\ 
a^m_{\pi\pi} = -S_{\pi\pi} &=& 0.74\pm 0.16;\nonumber\\ 
\acp(\pi^+\pi^0) &=& -0.07\pm 0.14;\nonumber\\ 
\acp(K^0\pi^+) &=& -0.02\pm 0.06;\nonumber\\
\acp(K^+\pi^0) &=& 0.00\pm 0.07;\nonumber\\
\acp(K^+\pi^-) &=& 0.095\pm 0.028;\nonumber\\
\acp(K^0\pi^0) &=& -0.03\pm 0.37;\nonumber\\
a^d_{K_S\pi^0} = C_{K_S\pi^0} &=& 0.40^{+0.27}_{-0.28}\pm 0.09; \nonumber\\
a^m_{K_S\pi^0} = -S_{K_S\pi^0} &=& 0.48^{+0.38}_{-0.47}\pm 0.28.
\end{eqnarray}

\begin{table}
\begin{tabular}{||c||c|c|c||}
\hline
Mode & CF & QCDF & PQCD\\
\hline
$\pi^+\pi^-$ & 9.0 - 12.0 & 3.5 - 15.0 & 5.5 - 9.0 \\
             & 19.7 - 21.6 & $-4$ - 21 & $-30$ - $-16$\\
$\pi^0\pi^0$ & 0.35 - 0.63 & 0 - 0.07 & 0.2 - 0.4 \\
             & $-42.2$ - 45.9 & $-100$ - 0 &     \\
$\pi^+\pi^0$ & 3.0 - 6.8 & 4.9 - 6.1 & 2.6 - 5.0\\
             & 0.0 - 0.1 & 0.0 & 0.0 \\
$K^+\pi^-$ & 14.0 - 18.0 & 3.0 - 24.5 & 13.0 - 18.6\\
             & $-10.6$ - $-3.7$ & $-13$ - $-6$ & 12.9 - 21.9\\
$K^0\pi^0$ $\ddag$ & 5.0 - 7.4 & 1.5 - 10.0 & 8.3 - 8.9\\
             & 24.4 - 36.5 & 0 - 7 &  \\
$K^+\pi^0$ & 9.4 - 12.0 & 2.0 - 16.0 & 7.6 - 11.0\\
             & $-9.2$ - $-2.6$ & $-10.6$ - 3.0 & 10.0 - 17.3 \\
$K^0\pi^+$ $\ddag$ & 14.0 - 22.0 & 5.5 - 26.0 & 13.7 - 19.7\\
             & $-1.5$ - $-1.3$ & $-1.75$ - 0 & 0.6 - 1.5\\
\hline
\end{tabular}
\caption{Predictions for BRs (first row, multiplied by $10^6$) and 
$\acp$s (multiplied by 100, and in our convention) of different nonleptonic 
modes. The numbers are taken from \cite{ali,beneke2,keum1,keum2}. The 
uncertainties are not treated equally; in particular, QCDF pedictions
take into account more sources of uncertainty than the other two. The
UT angle $\gamma$ is fixed to be in the first quadrant. For the modes indicated
by $\ddag$, the CP asymmetry is measured with a $K_S$ in the final state.} 
\end{table}

The theoretical predictions based on different models are given in Table 1. 
The `large' BRs ({\em i.e.}, except that of $B\to\pi^0\pi^0$) more or less
agree with each other; there is a nontrivial overlap region. Also to be
noted that only the BR of $B\to\pi^0\pi^0$ is {\em significantly} off the mark.
The predictions for $\acp$s differ, mainly because an enhancement of
penguins and annihilation topologies in the PQCD model. However, even taking
into account all the theoretical uncertainties,
some of the experimental resuls do not tally with the predictions.
The error bars are large, and ultimately the
difference might go away, but this is high time to be prepared for any 
indirect signal of NP.

Let us first discuss the theory of $\bpipi$.
We follow the standard convention of writing the amplitudes \cite{buras1}
({\em i.e.}, putting the top-mediated penguins together with up- and
charm-mediated penguins by using the unitarity relationship
$V_{tb}V_{td/s}^* = -V_{ub}V_{ud/s}^* - V_{cb}V_{cd/s}^*$)  
except that the EWPs, both color-allowed and color-suppressed, are written
separately. This will help us to understand the EWP dynamics of the
$\pi\pi$ sector. The amplitudes are
\begin{eqnarray}
A(\bar B\to\pi^+\pi^-)&=&|T_c|e^{-i\gamma} + |P_c| e^{i(\d_P-\d_T)}\nonumber\\
&{}& +|P^C_{EW}|e^{i(\d_{EC}-\d_T)} {\cal G}, \nonumber\\
\sqrt{2} A(\bar B\to\pi^0\pi^0) &=&
-|C|e^{-i\gamma} + |P_c| e^{i(\d_P-\d_C)}\nonumber\\
&{}&-{1\over 2} |P^C_{EW}|e^{i(\d_{EC}-\d_C)} {\cal G}\nonumber\\
&{}&+{3\over 2} |P_{EW}|e^{i(\d_{E}-\d_C)} {\cal G},
\nonumber\\
\sqrt{2} A(B^-\to\pi^-\pi^0) &=&
|T_c|e^{-i\gamma} +
|C|e^{-i\gamma}e^{i(\d_C-\d_T)}\nonumber\\
&{}&+{3\over 2} |P^C_{EW}|e^{i(\d_{EC}-\d_T)} {\cal G}\nonumber\\
&{}&-{3\over 2} |P_{EW}|e^{i(\d_{E}-\d_T)} {\cal G},\nonumber\\
&{}&
\end{eqnarray}
where ${\cal G} = 1 + R_b \exp(-i\gamma)$, and
$R_b = |\lambda_u/\lambda_c| = 0.37\pm 0.04$, with $\lambda_i=V_{ib}V_{id}^*$.
Conventionally, $T_c$ and $C$ contain, apart from the tree-level
amplitudes, those parts of the strong penguin that are proportional to
$e^{-i\gamma}$. For simplicity (and also guided by intuitive arguments) 
we take only the top-mediated EWP amplitudes; they are proportional to
$\lambda_t$, and can be written in terms of $\lambda_u$ and $\lambda_c$
using the unitarity argument. 
Note that we have extracted the strong phase associated with the leading
term, since only the difference of strong phases is a measurable
quantity. The annihilation contributions have been neglected, but can
be suitably dumped with
$T_c$ or $P_c$. To calculate the respective BRs, one has to remember that
$B^+$ lives longer than $B^0$ by a factor of $1.086\pm 0.017$ \cite{pdg2004}.  

Due to our slightly different valence quark convention,
the amplitude sum rule looks like
\begin{equation}
A(\pi^+\pi^-) - \sqrt{2} A(\pi^0\pi^0) = \sqrt{2} A (\pi^+\pi^0).
\end{equation}
This convention, of course, yields identical results to that of, say,
\cite{gronau}.

Thus, there are ten free parameters; five amplitudes, four independent
strong phase differences and the weak phase $\gamma$. The observables
are far less in number: only six, namely, three BRs and three CP asymmetries.
One therefore cannot say in a model-independent way that the EWPs are
going to be large. However, if one neglects the EWPs, the number of free
parameters gets reduced by four (two amplitudes and two strong phases),
so one can extract a unique solution \cite{buras1}. We, on the other hand,
will take the model predictions seriously. For example, we use the PQCD
prediction $R_c=|P_c/T_c|=0.23^{+0.07}_{-0.05}$ and 
$-41^\circ < \d \equiv \d_P-\d_T < -32^\circ$. We also use the CKM
fit value for $\gamma$: $50^\circ < \gamma < 72^\circ$ at 68\%
CL. This gives us three more theoretical inputs.

Next we come to the $\bpik$ modes. One can invoke the flavor SU(3) and
relate the $\bpipi$ amplitudes to the $\bpik$ ones. The symmetry is 
supposed to be broken only by the corresponding CKM factors and the 
ratio of the meson decay constants $f_K/f_\pi$. This in turn implies that
one takes the chiral enhancement factors coming from $(V-A)\otimes (V+A)$ 
type operators ($O_5$-$O_8$ in the standard
literature) to be the same in both cases.
Let us follow a different approach. We start with the assumption that
the amplitudes are independent of each other; later we will see whether
there is any trace of the flavor symmetry. If one can relate the amplitudes
by SU(3), that will act only as a further boost to our analysis.
In fact, the PQCD relationship
between the amplitudes $T_c$ and $P_c$, and the analogous one for $b\to s$
transitions, $|T'_c/P'_c|= 0.201\pm 0.037$, is consistent with the SU(3)
expectation
\begin{eqnarray}  
T'_c &=& T_c {f_K\over f_\pi} {V_{us}\over V_{ud}} = 0.28 T_c;\nonumber\\
P'_c &=& P_c {f_K\over f_\pi} {V_{cs}\over V_{cd}} = -5.3 P_c,
\end{eqnarray}
where primes indicate $b\to s $ transitions. Note that PQCD also predicts
the strong phase difference between $P_c'$ and $T_c'$ to be about $156^\circ$
\cite{keum2}.
We, however, will not use that as an input; rather, we will show that the
allowed solutions, obtained just from fitting the data, are entirely
consistent with the PQCD prediction.

We use the Wolfenstein representation of the CKM matrix, and define
$\bl = \l/(1-\l^2/2) \approx 0.230$, and
\begin{equation}
{R_b}'= R_b (d\to s) = R_b {V_{us}\over V_{ud}} {V_{cd}\over V_{cs}}
=-\bl^2 R_b.
\end{equation}
Remember that $V_{cd}$ is negative, hence the minus sign. We assume 
that there is no effect of NP or unknown SM dynamics in $\bbbar$ mixing,
and the phase from the $\bbbar$ box is $\sin(2\beta)$, which has been
directly measured as well as fitted. We use the CKMfitter fit result
$\sin(2\beta) = 0.739\pm 0.048$. It was discussed earlier what happens if
one relaxes this assumption and allows NP effects in the mixing. 

The amplitudes can be written in an analogous way to that of the $\bpipi$ 
ones. There is, howver, one important difference. The modes 
$B^+\to K^0\pi^+ $ and $B\to K^0\pi^0$ have no tree-level contributions.
The strong penguin, which after rearrangement was absorbed into $T_c$,
now gives a different amplitude $P'_{ct} \equiv P'_c-P'_t$ 
with a different strong phase 
$\d'_{CT}$ (we use primes to indicate $b\to s$ transitions). 

\begin{widetext}
\begin{eqnarray}
A(\bar{B}\to K^-\pi^+) &=& |T_c'|e^{-i\gamma} + |P_c'|e^{i(\d'_P-\d'_T)}
+|{P^C_{EW}}'|e^{i(\d'_{EC}-\d'_T)}{\cal G}',\nonumber\\
A(B^-\to \bar{K^0}\pi^-) &=& -|P_c'| - |P_{ct}'|e^{-i\gamma}e^{i(\d'_{CT}-
\d'_P)}
+{1\over 2}|{P^C_{EW}}'|e^{i(\d'_{EC}-\d'_P)}{\cal G}',\nonumber\\
\sqrt{2} A(B^-\to K^-\pi^0) &=& |T_c'|e^{-i\gamma} + |P_c'|e^{i(\d'_P-\d'_T)}
+|{P^C_{EW}}'|e^{i(\d'_{EC}-\d'_T)}{\cal G}' + \nonumber\\
&{}& \left[\bl |\tilde C|e^{-i\gamma}e^{i(\d_{\tilde C}-\d_T)}
+{1\over\bl} {3\over 2}|P_{EW}| e^{i(\d_E-\d_T)}{\cal G}'\right]\nonumber\\
\sqrt{2} A(\bar{B}\to \bar{K^0}\pi^0) &=&
|P_c'| + |P_{ct}'|e^{-i\gamma}e^{i(\d'_{CT}-\d'_P)} - {1\over 2}
|{P^C_{EW}}'|e^{i(\d'_{EC}-\d'_P)}{\cal G}' + \nonumber\\
&{}& \left[\bl |\tilde C|e^{-i\gamma}e^{i(\d_{\tilde C}-\d_P)}
+{1\over\bl} {3\over 2}|P_{EW}| e^{i(\d_E-\d_P)}{\cal G}'\right].
\end{eqnarray}
\end{widetext}

In the above expressions, ${\cal G}' = 1 + R_b'e^{-i\gamma}$,
and $\tilde C$ is the same as $C$ (in $\bpipi$) except that there is
no strong penguin (this difference is non-negligible). In the last two
amplitudes, there is a part where the $\pi$ is emitted from the $B$
meson; this, expectedly, is very similar in structure to that of $\bpipi$.
In these terms, 
the relative sign between $\bl$ and $\bl^{-1}$ terms is positive
whereas in the analogous expression for $\bpipi$, that is negative; this
is because $V_{cd}$ is negative. $P_{EW}$ is, of course, the same as that
in $B\to\pi\pi$. Thus, there are five new amplitudes and four strong
phases. 

\section{Analysis: $\bpipi$}

The main emphasis of the $\bpipi$ analysis is to extract, unambiguously,
a signature for large EWP amplitudes. In the SM, we expect $T_c$ to be
the largest amplitude, while $P_c$ and $C$ are smaller than $T_c$. In fact,
models like PQCD predict $|P_c/T_c|\sim 0.25$, while the predictions from
other models are generally below 10\%. There is no such prediction 
available for $C$, but one can expect $|C/T_c| < 1$. The EWP 
amplitudes are expected to be highly suppressed compared to these. The 
color suppression further diminishes $P_{EW}^C$ compared to $P_{EW}$. 

For those readers who do not want to go through the nitty-gritties of
the analysis, let us state our conclusions right here. We will see that
the amplitudes $T_c$ and $C$ turn out to be as expected ($P_c$ is related
to $T_c$), and though the color-allowed EWP $P_{EW}$ may happen to be large,
there are a number of solutions (we have more unknowns than equations, so
there is a large number of possible solutions) where this turns out to be
small, almost as small as expected from the theory. Later, we will see that
only those small values are allowed by the $\bpik$ data. The surprise comes
from the color-suppressed EWP amplitude $P^C_{EW}$; it turns out to be large,
larger than $P_c$, and almost as large as $T_c$! There are other interesting
features of the analysis; we discuss them at the end of this section.

We start by treating all the amplitudes as free parameters, with the exception
of $P_c$. The latter is varied over its entire range allowed by the
theoretical uncertainties of $|P_c/T_c|$. 
The other four amplitudes are varied over the
range 0-$6\times 10^{-8}$. We will see later that this is a rather conservative
range. All the strong phase differences are varied over the range 0-$2\pi$,
except $\d_P-\d_T$, which is again constrained by the theory. The weak angle
$\gamma$ is varied between $50^\circ$ and $72^\circ$.

From the solutions that satisfy the experimental constraints, we get the
following ranges for the amplitudes:
\begin{eqnarray}
2.5 < &T_c\times 10^8& < 4.1;\nonumber\\
0.4 < &P_c\times 10^8& < 1.2;\nonumber\\
0 < &C\times 10^8& < 3.8;\nonumber\\
0 < &P_{EW}\times 10^8& < 2.5;\nonumber\\
1.0 < &P^C_{EW}\times 10^8& < 3.7.
\end{eqnarray}
Note that apart from the last line, none of these are unexpected. One 
may object that $C$ is supposed to be smaller than $T_c$; if we take that
constraint into account nothing changes except the upper limit of $C$;
it becomes $3.5\times 10^{-8}$. However, there is a concenration of
solution points in the $C-P_{EW}$ plane near the origin (see Fig. 1). 
This vindicates the conjecture that both these amplitudes may be small.
Later we will see from the $\bpik$ channels that the upper limit of 
$P_{EW}$ is smaller than the range predicted here by about 40\%. 
\begin{figure}[htbp]
\vspace{-10pt}
\centerline{\hspace{-3.3mm}
\rotatebox{-90}{\epsfxsize=6cm\epsfbox{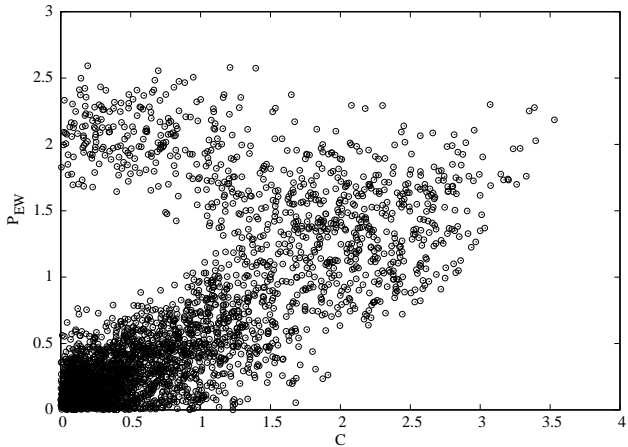}}}
\caption{$C$ versus $P_{EW}$ for $\bpipi$. Note the concentration of
solutions near the origin, and the excluded region for $P_{EW}$ for
low $C$. The amplitudes have been multiplied by $10^8$.}
\end{figure}

Why $P^C_{EW}$ is so large? The main culprit is the $B\to\pi^0\pi^0$ BR.
This, along with the facts that the other two BRs are more or less in the
expected ballpark, while the direct CP asymmetry in $B\to\pi^+\pi^-$ 
is large ({\em i.e.}, there must be a significant interfering amplitude)
generates a large $P^C_{EW}$. See Fig. 2 to have an idea of the solution
points for $P^C_{EW}$; in fact, if we claim that $P_{EW}$ is small, the range 
for $P^C_{EW}$ becomes 2-3$\times 10^{-8}$. This is a large amplitude
by any standard, and cannot be accomodated in any theoretical model. 

The strong phase differences do not show any interesting pattern, except,
again, for $\d_{EC}-\d_T$. Varied between 0 and $2\pi$, solutions come out
only for the range $174^\circ < \d_{EC}-\d_T < 190^\circ$. This shows
a large destructive interference between $P^C_{EW}$ and $T_c$, which is
necessary to reproduce the BRs of $B\to\pi^+\pi^-$ and $B^+\to\pi^+\pi^0$
in the right ballpark.
\begin{figure}[htbp]
\vspace{-10pt}
\centerline{\hspace{-3.3mm}
\rotatebox{-90}{\epsfxsize=6cm\epsfbox{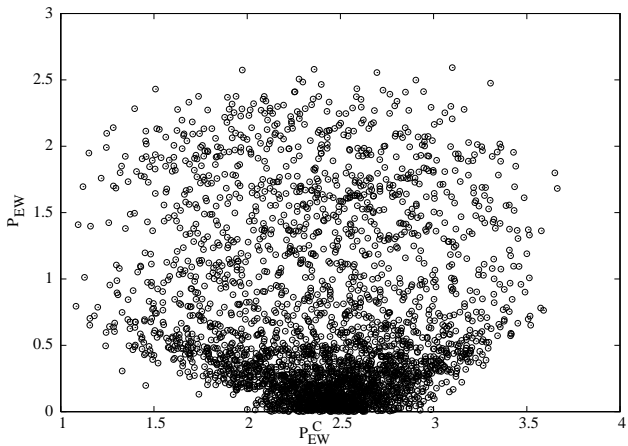}}}
\caption{$P_{EW}$ versus $P^C_{EW}$ for $\bpipi$. Note the large nonzero
allowed values for the latter, and the concentration of
solutions for small $P_{EW}$.
The amplitudes have been multiplied by $10^8$.}
\end{figure}

Fig. 3 shows the allowed parameter space of $a^d_{\pi\pi}$ versus $a^m_{\pi
\pi}$; this shows a concentration of solutions near the physical
boundary $|a^d_{\pi\pi}|^2 + |a^m_{\pi\pi}|^2  \leq 1$. 

\begin{figure}[htbp]
\vspace{-10pt}
\centerline{\hspace{-3.3mm}
\rotatebox{-90}{\epsfxsize=6cm\epsfbox{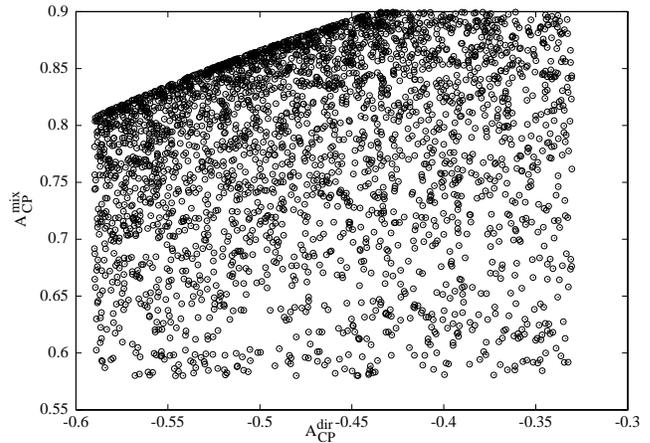}}}
\caption{Direct and mixing-induced CP asymmeries in $B\to\pi^+\pi^-$.
The circular arc that bounds the parameter space at the upper left hand
corner corresponds to $|a^d_{\pi\pi}|^2 + |a^m_{\pi\pi}|^2  = 1$.}
\end{figure}
 
Let us emphasize again at this point 
that this treatment is valid if and only if
the new physics, if any, contributes to the decay but not to the $\bbbar$
mixing. Most of the new physics models are of this type; however, there 
are exceptions. For example, four-generation fermion models contribute
mostly in mixing. Supersymmetry with R-parity violation may contribute to 
only mixing, only decay, or both, depending upon the nonzero couplings chosen.
Generically, if NP contributes in $\bbbar$ mixing, the following modifications
have to be taken into account (an example with R-parity violation has been
exhaustively discussed in \cite{bdk2}):

\begin{itemize}

\item $\Delta M_d$ is affected by NP. Consequently, $V_{td}$, which is 
determined from the measurement of $\Delta M_d$,
becomes a free parameter, except for constraints coming from
the unitarity of the CKM matrix. Obviously, for models with more than three
quark generations, this is a very loose constraint. Fortunately, the precise 
value of $V_{td}$ is not that important in our analysis.

\item What is more important is the value of $\sin(2\beta)$. There is a SM
value of $\beta$, which let us call $\beta_{SM}$. We have explicitly
assumed that the CP asymmetry in, say, $B\to\psiks$, measures $\beta_{SM}$.
If there is NP in mixing, it will measure $\beta$ which should be 
different from $\beta_{SM}$. Note that this happens even if the NP mixing
amplitude is real. The value of $\beta$ is important in the analysis of
CP asymmetries in the $B\to\pi^+\pi^-$ channel. To get a feeling of 
$\beta_{SM}$, one should perform the usual CKM fit for the vertex $(\rho,
\eta)$ of the UT without the $\Delta M_d$ constraint. Again, this is
not a foolproof prescription; one implicitly assumes that NP does not 
contribute to, say, $\varepsilon'/\varepsilon$. 
\end{itemize} 

Even considering all this, it is hard to see how a NP that contributes
only to mixing but not to decay can explain the $\bpipi$ results. It can affect
the fit for $\gamma$, but cannot generate a large BR for $B\to\pi^0\pi^0$ 
or a large direct CP asymmetry for $B\to\pi^+\pi^-$.

\section{Analysis: $\bpik$}

The analysis for $\bpik$ follows that of $\bpipi$. There are five new
amplitudes, apart from $P_{EW}$, and hence four new strong phase
differences. We scan this ten-dimensional parameter space for possible
solutions. Note that $R_b'$ is a small number and hence for all
practical purpose ${\cal G}'$ can be approximated by unity (but for
our analysis we keep it as it is). Also note that these ten parameters
are not really all free; the allowed range for $P_{EW}$, as well
as the weak phase $\gamma$, are determined
from the $\bpipi$ data, and we use the PQCD result $|T'_c/P'_c|= 0.201\pm 
0.037$. We also equate the strong phases of $P_c'$ and $P_{ct}'$. 

We obtain the following ranges for the amplitudes:
\begin{eqnarray}
0.2 < &T_c'\times 10^8& < 1.15;\nonumber\\
0.8 < &P_c'\times 10^8& < 4.8;\nonumber\\
0 < &P_{ct}'\times 10^8& < 3.2;\nonumber\\
0 < &\tilde C\times 10^8& < 3.8;\nonumber\\
0 < &P_{EW}\times 10^8& < 1.55;\nonumber\\
0 < &{P^C_{EW}}'\times 10^8& < 6.3.
\end{eqnarray}

There are two points to note right here. First, $\tilde C$ could in principle
have a bigger range. We constrained it to be smaller than $C$ of $\bpipi$
modes. Second, both the electroweak penguin amplitudes can be close to zero;
we refer to Figure 4 for a feeling of the solutions.
Thus, {\em there is no apparent EWP anomaly in the data}. We have explicitly
checked that solutions are possible even if one puts by hand both the
EWP amplitudes equal to zero. 

All the strong phase differences have been varied between 0 and $2\pi$.
Solutions are allowed for the entire range, unlike $\bpipi$, where 
we found a rather constrained fit $174^\circ < \d_{EC}-\d_T < 190^\circ$.
However, from Figures 5 and 6, there appears to be an interesting
correlation between the amplitudes and the strong phases.

\begin{figure}[htbp]
\vspace{-10pt}
\centerline{\hspace{-3.3mm}
\rotatebox{-90}{\epsfxsize=6cm\epsfbox{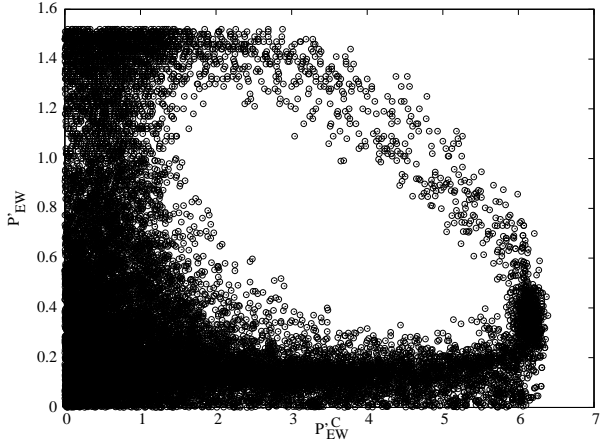}}}
\caption{Allowed solutions for color-allowed and color-suppressed EWP
amplitudes, showing that there is no apparent EWP anomaly in $\bpik$
data.
The amplitudes have been multiplied by $10^8$.}
\end{figure}

\begin{figure}[htbp]
\vspace{-10pt}
\centerline{\hspace{-3.3mm}
\rotatebox{-90}{\epsfxsize=6cm\epsfbox{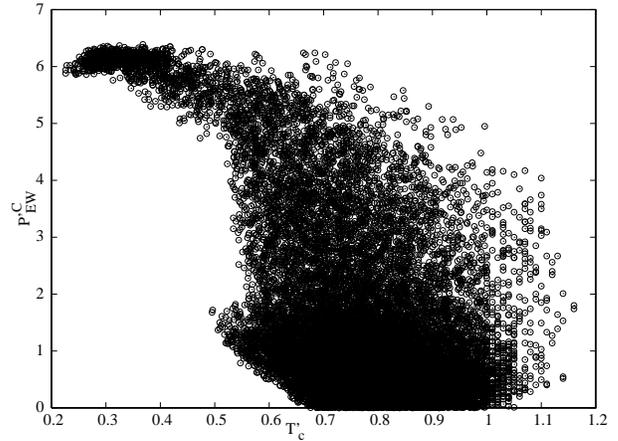}}}
\caption{
$T_c'$ versus ${P^C_{EW}}'$ for $\bpik$. Note how small
${P^C_{EW}}'$ solutions necessarily corespond to large $T_c'$
which is consistent with SU(3) relationship.
The amplitudes have been multiplied by $10^8$.}
\end{figure}

From Figure 5, it appears that small ${P^C_{EW}}'$ solutions are admitted
only if $T_c'$ is large: $|T_c'| > 0.55\times 10^{-8}$. For smaller values
of $T_c'$, ${P^C_{EW}}'$ must be large. This is intuitively clear: after
all, one of them should be large enough to generate the required BRs.
One could have done this taking $T_c'$ and $P_c'$ to be free parameters;
but we wish to follow the theoretical prediction, which is also
consistent with an SU(3) flavor symmetry, as far as possible. 

\begin{figure}[htbp]
\vspace{-10pt}
\centerline{\hspace{-3.3mm}
\rotatebox{-90}{\epsfxsize=6cm\epsfbox{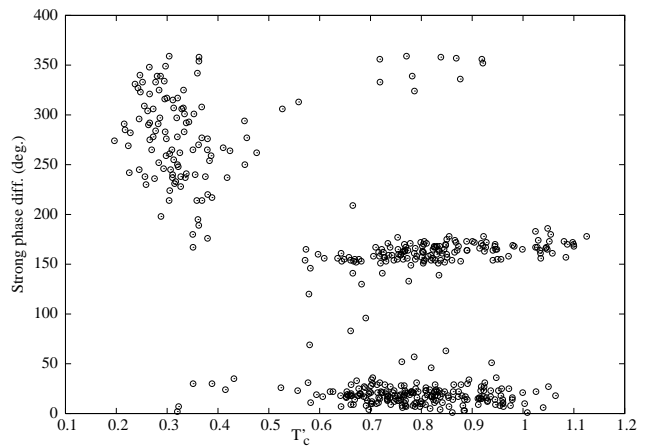}}}
\caption{
Allowed solutions for the strong phase difference $\d'_P-\d'_T$
as a function of $T_c'$, multiplied by $10^8$.}
\end{figure}

Figure 6 shows the solutions for the strong phase difference $\delta'_P
-\delta'_T$ as a function of $T_c'$. Note the band around the theoretical
prediction of $156^\circ$ from the PQCD model, which appears only for large
$T_c'$ and hence small EWPs. This shows that there is nothing to
worry about the EWP sector; but one can accomodate large values, which
may come from NP models. 

Thus, the SU(3) flavor symmetry is there, as far as the magnitudes of
$T_c^{(')}$, $P_c^{(')}$ and $P_{EW}^{(')}$ are concerned. However,
the strong phase differences are not 
necessarily small. If one demands a correlation between both the amplitudes
and the strong phases for $\bpipi$ and $\bpik$, one must invoke large
EWP contribution. But such a contribution is there for $\bpipi$; there is
no a priori reason why it should not appear for $\bpik$, and thus help
to salvage the flavor symmetry.  

\section{Summary and Conclusions}

We have juggled with a number of amplitudes and strong phases. The equations
being horrendously coupled, we do not even envisage to attempt an analytic
solution. Those solutions necessarily involve some simplification and run
the risk of missing the point. Rather, we solved these equations numerically,
and the equations being more in number than constraints (and remember that
these constraints often have large error bars), found a range of
allowed values for the free parameters. It is indeed interesting that one
can form some idea about the nature of the solutions.

We assume that the theoretical models can catch the essence of these
nonleptonic transitions successfully. There are points where they differ;
for example, the way how one takes care of end-point suppresions, higher-twist
corrections, or annihilation topologies. Fortunately, such subtleties 
are not really crucial for our analysis. What is important is the relative
strength of the strong penguin amplitudes with respect to the tree
amplitude. PQCD, by virtue of a dynamical enhancement mechanism, predicts 
a larger $|P_c/T_c|$ than other models like CF or QCDF. Since our strategy
is to extract the minimal set that does not fit in these model predictions,
we take PQCD as our working model. If one takes, for example, QCDF, one
needs a larger color-suppressed EWP to explain the $\bpipi$ data. PQCD
also helps to fit $\gamma$ in the first quadrant, which is compatible with
the CKM fit.

We found that the explanation of the $\bpipi$ data needs a large 
color-suppressed EWP contribution (the color-allowed EWP may be large, but
need not be). This is entirely due to the large BR of $B\to\pi^0\pi^0$.
If this data would have been in the expected ballpark, the large direct
CP asymmetry in $B\to\pi^+\pi^-$ could be explained by a much smaller
EWP contribution. Such a large EWP contribution also means that there
should be something worth watching in $B\to\rho\gamma$.

The situation is not so drastic for $\bpik$. Even if one takes the SU(3)
predictions of $T_c'$ and $P_c'$ seriously, there are solutions with
vanishingly small EWP amplitudes. Of course, there is a price to pay: the
other two amplitudes $\tilde C$ and $P_{ct}'$ become essentially free,
except for the constraints that $|\tilde C| < |C|$ and $\delta'_P
=\delta'_{CT}$. We have tried to see what happens to the fit if all
amplitudes and strong phases are taken from some model (we used CF,
because that predicts all the amplitudes and also the strong phases 
coming from the imaginary parts of the respective Wilson coefficients),
and there one indeed requires a sizable EWP contribution!

So, what is the lesson from the $\bpik$ analysis? Probably that it is yet
early to say that there is any definite evidence for large EWP amplitudes,
but they can be easily accomodated, as is evident from our figures. 
All the penguins are expected to be enhanced by about a factor of 5
when going from $\bpipi$ to $\bpik$. This trend can be accomodated
easily for the color-allowed EWPs, but appears to be very hard for 
the color-suppressed ones. Thus, even if the color-allowed EWPs can
be explained within the present models, the color-suppressed one poses
a serious challenge.

Is the situation so bizarre because we do not understand the dynamics
of penguins, or to be more general, that of low-energy QCD? No one knows,
but from intuitive arguments it is very hard to establish such a large
EWP amplitude compared to the strong penguins. One should not also
forget the possibility that the experimental data may change over the years,
though such an assumption does not help the analysis. 

Is the solution to be found in physics beyond the SM? Again, it is early
days, but one may be hopeful, more so because this may provide an indirect
evidence for such NP before the Large Hadron Collider (LHC) comes into
operation. NP models generate new operators in the effective Hamiltonian,
or modify the Wilson coefficients for the existing operators (an example is
the Universal Extra Dimension model of Appelquist {\em et al} \cite{ued}). 
If the interactions are flavor-specific, they will appear as a modification
to the tree or the EWP amplitudes. These models include exotic quarks, new
$Z'$ gauge bosons, R-parity conserving and violating supersymmetry and so
on. 

A couple of cautions here. First, one needs the exact structure of a model
to be more specific than the present analysis. What one can do is to eliminate
models: for example, supersymmetry with flavor alignment will not generate
such a large FCNC amplitude. The exact weightage of new operators will also
depend on the particular model chosen. Second, this analysis has to be
slightly modified if there is new physics in $\bbbar$ mixing. This has been
discussed in Section 3.

Where should one get confirmatory signals for such NP effects? Again, that 
depends on the specific model. We have already mentioned $B\to\rho\gamma$;
another place would be the semileptonic $b\to d$ decays. For nonleptonic
decays, one should carefully analyse the $B\to\rho\pi$, $B\to\rho\rho$, $B_s
\to\pi K$, and $B_s\to K K$ channels. An analysis of angular correlations
in $B\to\rho\rho$ \cite{sinha} should be useful.

\begin{acknowledgments}
S.N. has been supported by a UGC research fellowship.
\end{acknowledgments}

\end{document}